\documentstyle[preprint,aps]{revtex}  

\newcommand{\ul}[1]{\underline{#1}}

\tighten       

\begin{document}
\draft

\title{Impurity bands and sequential resonant tunneling
in the presence of terahertz fields}
\author{Andreas Wacker and Antti-Pekka Jauho}
\address{Mikroelektronik Centret,
Danmarks Tekniske Universitet, DK-2800 Lyngby, Denmark}
\author{Stefan Zeuner and S.~James Allen}
\address{Center for
Terahertz Science and Technology, University of California
at Santa Barbara, Santa Barbara, California 93106 }

\date{email: wacker@mic.dtu.dk}
\maketitle

\begin{abstract}
A theoretical and experimental study of
transport in a low doped multiple quantum well structure shows
that impurity bands 
are essential in understanding the electronic 
transport both with and without terahertz irradiation.  
A  full self-contained model,
which involves only the nominal parameters of the sample,
agrees quantitatively with a wide range of experiments.
\end{abstract}
\narrowtext

Perpendicular charge transport in biased superlattices displays a rich variety
of physical phenomena: negative differential resistance \cite{ESA70},
Bloch oscillations\cite{WAS93}, dynamic localization\cite{Zithol}, 
electric field domain formation\cite{Zitdom}, 
and photon-assisted transport\cite{KEA95a,KEA95b} are but a
few recently studied topics.  In
the case of weakly coupled multiple quantum well structures
the perpendicular transport
is dominated by resonant transitions between
the energy levels in neighboring wells.
In real systems these resonances are never sharp: they are broadened
by various scattering mechanisms due to, e.g., impurities,
interface roughness, phonons, or interactions between
the particles themselves (see, e.g., Ref.~\cite{MUR95}).
For samples with low doping impurity scattering may lead to the appearance
of impurity bands \cite{Zitser}.  These additional
narrow bands cannot be described by simple Lorentzian spectral
functions thus necessitating a more detailed microscopic model.
In the present paper we develop a fully microscopic model for this 
situation.
Our calculations yield a strong temperature dependence of the current-field
relation associated with the occurrence of two different peaks at low fields.
The implications of these features are compared 
to both previously published and new experimental results.

\begin{figure}
\vspace*{11.0cm}\includegraphics{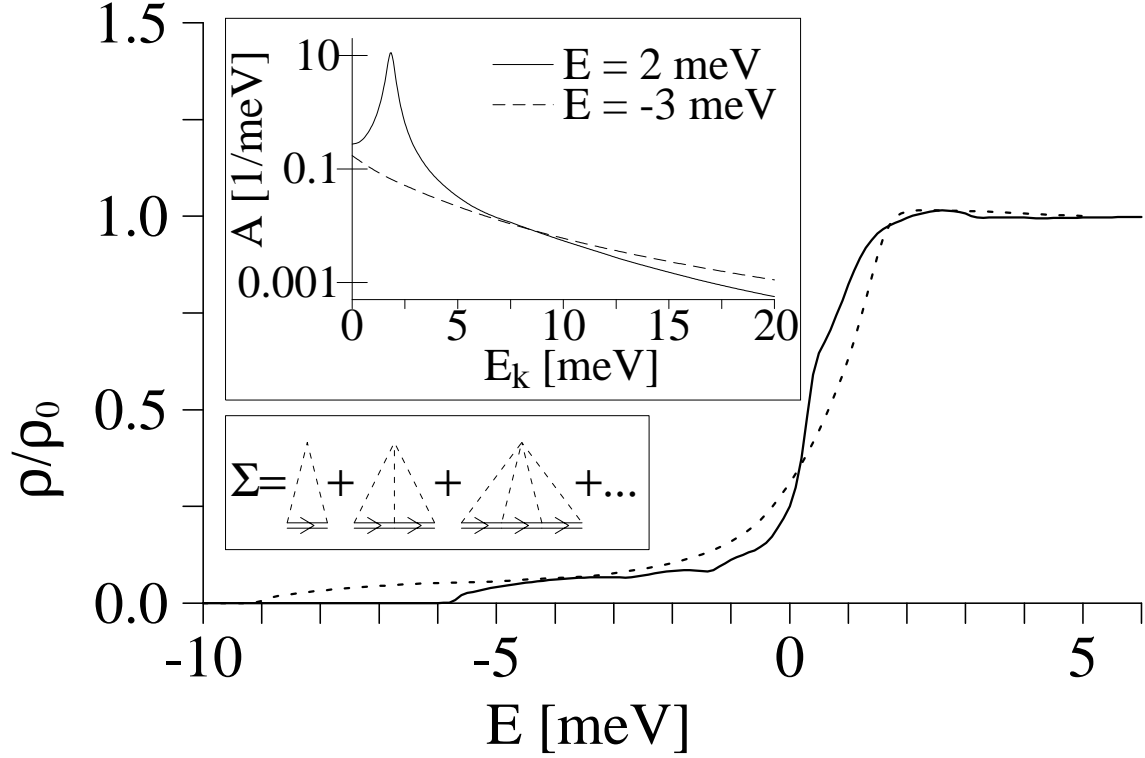}
\caption[a]{Calculated density of states in units of the 2D 
free carrier density $\rho_0$
using  screening due to a free electron gas (full line)
and reduced screening (dashed line).
The upper inset shows the spectral function $A(E_k,E)$
for different energies $E$. The lower inset depicts  diagrammatically the
self-consistent single-site approximation.}
\label{Figdichte}
\end{figure}

For weakly coupled quantum wells  
the current  from the $\nu^{th}$ level in well
$n$ to the $\mu^{th}$ level in well $n+1$ is given by \cite{MAH90}
\begin{eqnarray}
I_{n\to n+1}^{\nu \to \mu}=2e\sum_{\ul{k}}
|H_{1}^{\mu,\nu}|^2
\int_{-\infty}^{\infty} \frac{dE}{2\pi \hbar}A_{n+1}^{\mu}(\ul{k},E+eFd)\nonumber \\
\times A_n^{\nu}(\ul{k},E)\left[n_F(E-\mu_n)-n_F(E+eFd-\mu_{n+1})\right]
\label{EqJ}\, .
\end{eqnarray}
Here $e$ is the electron charge, $\mu_n$ is the electro-chemical
potential in well $n$, $n_F({\cal E})=1/[1+\exp({\cal E}/k_BT_e)]$,
and $T_e$ is the electron temperature.
$Fd$ is the voltage drop per period $d$.
The transition matrix elements
$H_{1}^{\mu,\nu}$ between adjacent wells are calculated using the Wannier
functions obtained from a Kronig-Penney
model \cite{sample}.
Here we restrict ourselves to resonant transitions where the tunneling
process conserves the wave vector $\ul{k}$ within the two-dimensional
electron-gas; nonresonant transitions (see Ref.~\cite{WACpb})
are of minor significance  for the sample considered here.
The microscopical description of the impurity scattering enters
through the spectral functions $A_n^{\nu}(\ul{k},E)$
which are evaluated in a self-consistent single-site
approximation
(shown diagrammatically in the lower inset of Fig.~\ref{Figdichte}) 
like in Ref.~\cite{Zitser}.
The homogeneous doping is modeled by 8 equally spaced 
$\delta$-doping layers per period.
The screening of the impurity interaction
is accounted for within the random phase approximation.
We restrict ourselves to the two lowest levels. For the upper level
we include a contribution to the imaginary part of the
self-energy due to optical phonon transitions to the ground level,
calculated as in \cite{FER89}.
We do not consider the spin-resolved electron-electron
interaction  leading to the splitting of the
impurity bands for a single impurity location, 
(the Mott transition, see e.g. \cite{SHK84})
because the uniform doping will lead to a smearing out
of these features.
The technical details of our calculation will be given elsewhere;
here we point out that the theoretical framework has recently \cite{WACpb}
been  successfully applied to high-doping samples, where impurity
bands play no role.


We next describe our numerical results.
The upper inset of Fig.~\ref{Figdichte} shows the spectral functions for
two different energies below ($E = - 3$ meV)
and above ($E = 2$ meV) the bottom of the (free electron)
band edge.
For $E = 2 $ meV the spectral function exhibits a 
characteristic Lorentzian-type
behavior  with a peak around $E_k=\hbar^2k^2/2m\approx 2$ meV.
This is the generic behavior of a quasiparticle with a finite
lifetime due to
scattering. In contrast to this behavior the spectral function is quite flat
for   $E=-3$ meV, which is a signature of an  impurity band\cite{Zitser}.
The density of states $\rho(E)=2/(2\pi)\sum_{\ul{k}}
A(\ul{k},E)$ for the lowest level is plotted in Fig.~\ref{Figdichte}
(full line). The impurity bands from different locations of
impurities overlap and  manifest themselves in a low density of
states in the range $E_{{\rm min}}=-5.8$ meV$<E<0$ below the band edge.
Importantly, for these low-doped samples
the Fermi level is  at $E_F=-2$ meV, i.e., {\em within} the
impurity band.  As we shall see below, this property has
far-reaching consequences.

\begin{figure}
\vspace*{9.2cm}\includegraphics{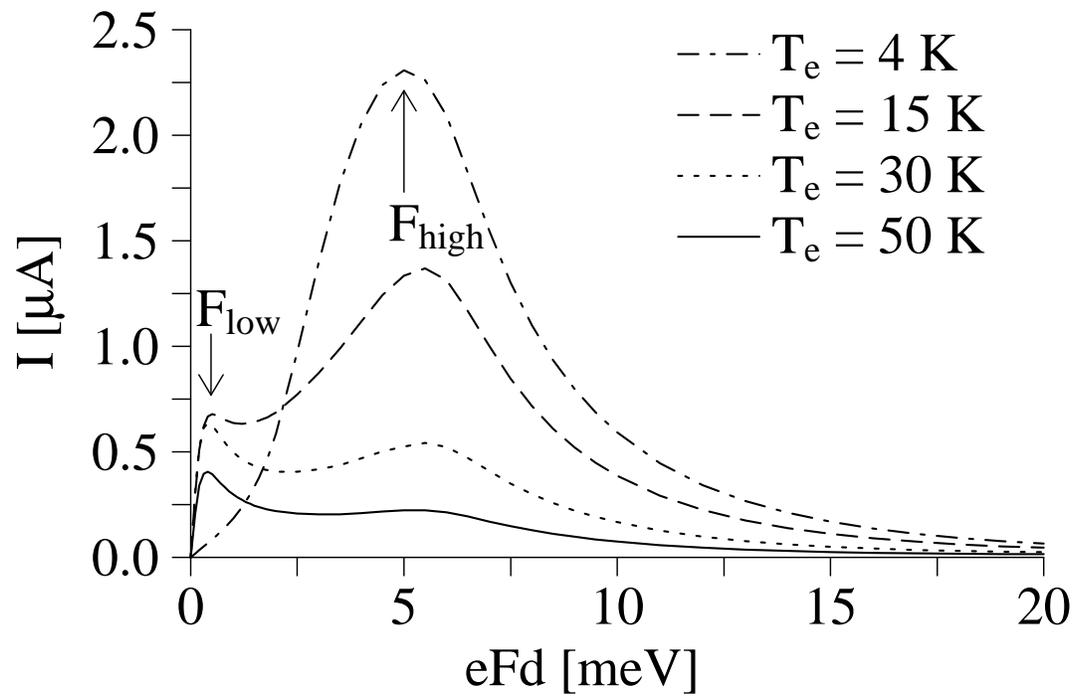}
\caption[a]{Current-field relation 
for different electron temperatures.}
\label{Fighom}
\end{figure}

We calculate the currents $I_{n\to n+1}(eFd)$ for
different electron temperatures $T_e$ by summing up the different
contributions from Eq.~(\ref{EqJ}).
Characteristic results are shown in Fig.~\ref{Fighom} for an
electron density $6\times 10^9$/cm$^2$ per well provided by the doping.
Here we focus in the low field
region where only tunneling between the lowest levels
in each well takes place.
For  $T_e = 4$ K we find a maximum  at $eF_{\rm high}d\approx 5$ meV.
This is due to tunneling from the impurity band to the free states.
The maximum occurs at the energy where the bottom of the impurity band
in one well is aligned with the band edge of the free electron states in the
neighboring  well, i.e., $eF_{\rm high}d\approx |E_{{\rm min}}|$.
An increasing temperature leads to a transfer of electrons
from the impurity band to the free-electron states and
consequently the current  at $eF_{\rm high}d$ decreases 
with increasing $T_e$. 
The density of states in the impurity band is much
lower than in the free-electron states, and hence
the majority of the electrons will be in the free-electron
states for  $k_BT_e\gtrsim |E_{{\rm min}}|$.
For the free-electron states the overlap between
the spectral functions (note that the inset of Fig.~\ref{Figdichte}
is plotted in log-scale)
from well $n$ and $n+1$ is very large at small fields.
This yields a strong enhancement in the low field
conductivity.  As the overlap of the
spectral functions becomes small  for voltage differences larger than their
broadening, the current {\em decreases} for  
$eFd>eF_{\rm low}d\approx 0.5$ meV.
This transition from a single-peaked to a
double-peaked current-field relation with increasing temperature
is essential in understanding the transport in low-doped
samples.

We can now resolve a recent puzzle of Ref.~\cite{ZEU96}:
there it was found that peaks
observed in the current-voltage characteristics (IV) under
terahertz irradiation 
can be interpreted consistently as photon-replicas
of a maximum in an \lq\lq instantaneous\rq\rq IV 
located at $U \approx 20 $ mV.
Yet, a direct measurement of the  IV without irradiation showed
no maximum at this bias.
A qualitative explanation, based on impurity bands, is
as follows.
For low electron temperatures and
without irradiation the maximum at $eF_{\rm high}d$
dominates the domain formation which sets in at voltages 
exceeding $U\approx NeF_{\rm high}d$ where $N=10$ is the number of wells. 
If the THz radiation is present
the electrons are excited from the impurity band into the free electron states
corresponding to a larger effective electron temperature.
Thus, the maximum at $U=NeF_{\rm low}d$ 
is dominant, and the photon replicas corresponding to this feature
are seen experimentally. 


While the above consideration explains qualitatively the findings
of Ref.~\cite{ZEU96},
it needs to be refined to agree quantitatively with the experiment:
the experimental photon-replica suggest 
$eF_{\rm low}^{\rm exp}d\approx 2$ meV 
which is four times the value from Fig.\ \ref{Fighom}.
The domain formation sets in at $U = 0.1$ V yielding
$eF_{\rm high}^{\rm exp}d\approx 10$ meV (i.e., twice larger
than $eF_{\rm high}d$ of Fig.\ \ref{Fighom}),
which seems to require  
$|E_{{\rm min}}|\sim 10$  meV.  This value follows from
our theory as well, as we now argue.
The screening of the impurity potential was treated
using a free electron density-of-states
$\rho_0= m/\pi\hbar^2$ yielding the polarizability
$\Pi(k)=\rho_0\left[1-\theta(k-2k_F)\sqrt{1-4(k_F/k)^2}\right]$.
Now $\Pi(0)$ is related to the {\em actual density of states} at 
$E_F\approx -2$ meV  which is
significantly lower than $\rho_0$.
Calculations within the Born-approximation
show that the $k$-dependence of the polarizability becomes weaker
and that $\Pi(0)$ decreases with increasing scattering \cite{AND82a,DAS83}.
In order to accommodate these trends we make the replacement
$\Pi(k)\to\Pi^*(k) = 0.07\rho_0$, given by the calculated
density of states of Fig.\ \ref{Figdichte}, and use this value
in further calculations.
The smaller polarizability
weakens the screening of the impurity potential and
therefore increases the binding energy of the impurities \cite{Zitser}.
The resulting density of states is shown in Fig.~\ref{Figdichte}
by a dashed line.
The  onset of the impurity band is now
$E_{{\rm min}}=-9.1$ meV and the current-field characteristics
look similar to Fig.~\ref{Fighom} but with $eF_{\rm low}d=2$ meV,
$eF_{\rm high}d=8.2$ meV in good agreement with the values deduced from
the experiment.

\begin{figure}
\vspace*{13cm}\includegraphics{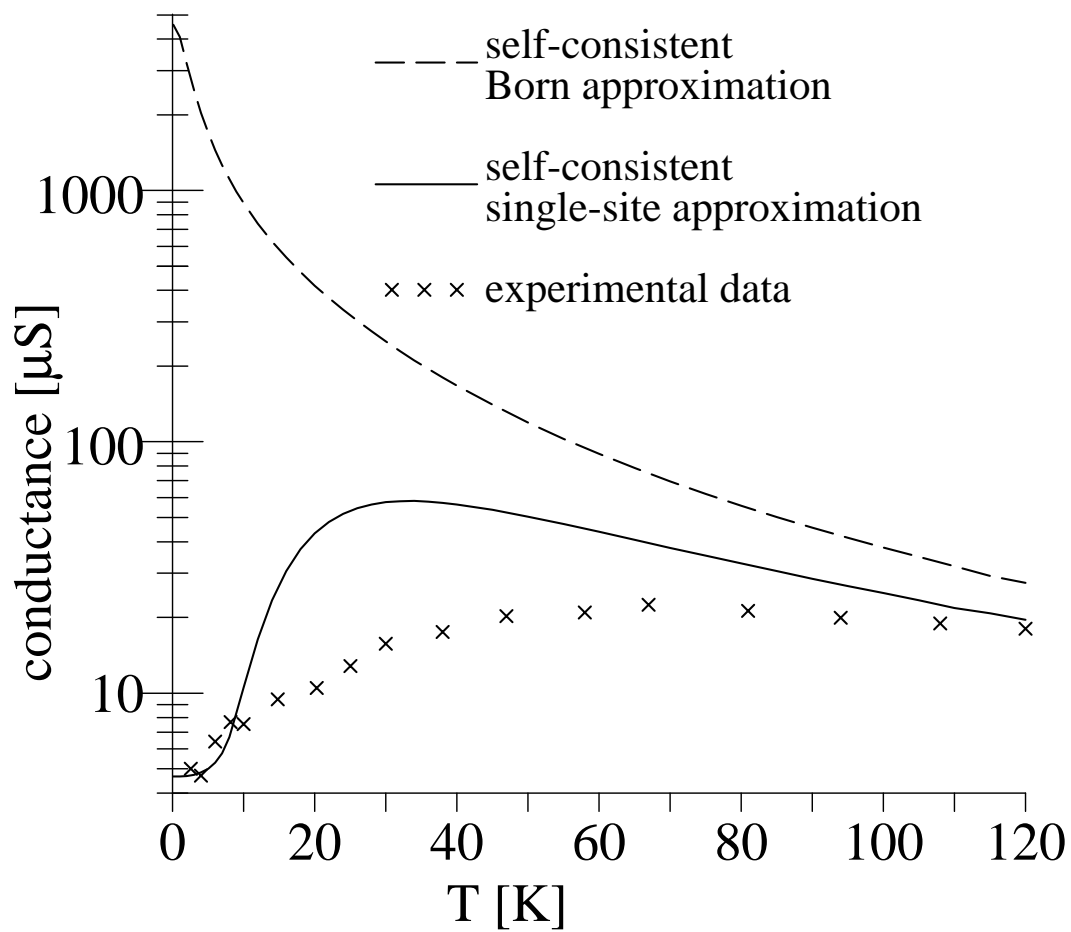}
\caption[a]{
Comparison of experimental
and theoretical  conductance at zero bias versus temperature.}
\label{Figmobil}
\end{figure}

As a further test of the theory,
we next consider the temperature dependence of the zero-bias conductance
$\sigma$.
Eq.~(\ref{EqJ}) yields
$\sigma(T)\propto\int dE\sum_{\ul{k}}A(\ul{k},E)^2\partial n_F(E)
/\partial E$, which
implies  $\sigma(T)\propto 1/ T$ for
$k_BT\gg E_F-E_{\rm min}$ and 
$\sigma(T) \to$ const. for $k_BT\ll E_F-E_{\rm min}$,
provided that $\sum_{\ul{k}}A(\ul{k},E)^2$ has no strong energy dependence.
This is the case if impurity scattering is treated within 
the self-consistent Born approximation where no impurity bands form. 
Then $\sigma$ is monotonously decreasing
in $T$ as shown in Fig.~\ref{Figmobil}.
However, a different scenario emerges if
the electrons occupy impurity bands for
low temperatures. Then  $\sigma$ is strongly suppressed due to 
the small values of $A(\ul{k},E)$ for $E<0$, see Fig.~\ref{Figdichte}.
As temperature is increased, more
electrons are excited to the free electron states, and $\sigma$ {\em increases}
with $T$ until the impurity bands are almost
empty at  $k_BT\sim |E_{{\rm min}}|$.  
This physical picture is 
confirmed by our measurements, shown in  Fig.~\ref{Figmobil}, together 
with our full calculation.
At low temperatures the agreement is quantitative,
while at intermediate $T$ the theory overestimates $\sigma$; 
this is most
likely due to additional scattering processes not included in our calculation,
or by the presence of a contact resistance which may limit the experimental
conductance.

\begin{figure}
\vspace*{9.2cm}\includegraphics{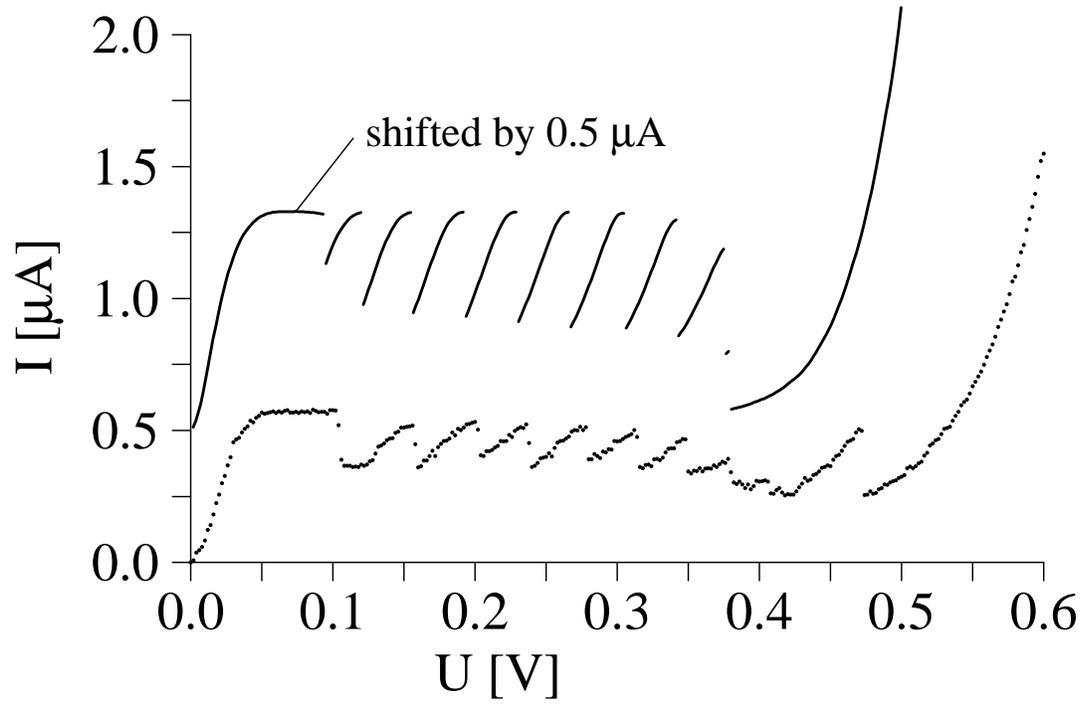}
\caption[a]{Experimental (dots) and
theoretical (full line)
current-voltage characteristics without irradiation.}
\label{Figdomaen}
\end{figure}

In order to calculate the full current-voltage characteristic
we must consider electron heating, i.e., estimate the
parameter $T_e$ in Eq.~(\ref{EqJ}).
With average voltage drop of
8 mV per well we find experimentally  $I=0.57$ $\mu$A without
irradiation. This gives a power dissipation
of $P=9$ pW per electron and a temperature
of $T_e\approx 47$ K using the standard
energy loss by optical phonons\cite{DAS92},
$\hbar \omega_{\rm LO}/\tau_{\rm LO}\exp(-\hbar \omega_{\rm LO}/kT_e)$
with $\tau_{\rm LO}\approx 80$ fs.
An increase of the energy loss with decreasing
electron density has been predicted due to plasmon-phonon 
coupling \cite{DAS92}.
Recently, $P=9$ pW at $T_e\approx 40$ K
for a sample with an electron density $6\times 10^{10}$/cm$^2$ per period
has been reported\cite{HILp}.
As our density is much smaller an even  lower electron temperature
is possible.
Experimentally, we find  that the
current-voltage characteristics exhibits a plateau for 
0.04 V$<U<$0.1 V  with $I=0.57$ $\mu$A 
(see Fig.~\ref{Figdomaen}). 
The current of this plateau
remains almost constant for lattice temperatures between 2 K and 30 K, but
drops for higher  temperatures.
This can be understood if the electron heating
prevents the electron temperature from cooling below 30 K. All these
facts indicate that
$T_e\approx 30$ K is a consistent choice for medium voltages.
For vanishing bias, on the other hand,
the electron temperature must approach  the lattice
temperature (4K).

The IV including
domain formation\cite{Details} is calculated as in Ref.~\cite{WACpb} and
is shown in Fig.~\ref{Figdomaen}.
We find excellent agreement with the data 
for the maximum current for the first peak
($I^{\rm peak}_{1,{\rm th}}=0.8\mu$A;
$I^{\rm peak}_{1,{\rm exp}}=0.58\mu$A) and the 
second peak at higher voltages
($I^{\rm peak}_{2,{\rm th}}=11\mu$A;
$I^{\rm peak}_{2,{\rm exp}}=13.6\mu$A, not shown in the figure).
The voltage where domain formation sets in is in good agreement with the
experimental value, but the domain branches occur at higher voltages in the 
experiment. This can be explained by the fact that once the high field
domain is formed it extends into the low doped receiving contact region
where a voltage drop of 0.1 V may occur.

\begin{figure}
\vspace*{18cm}
\includegraphics{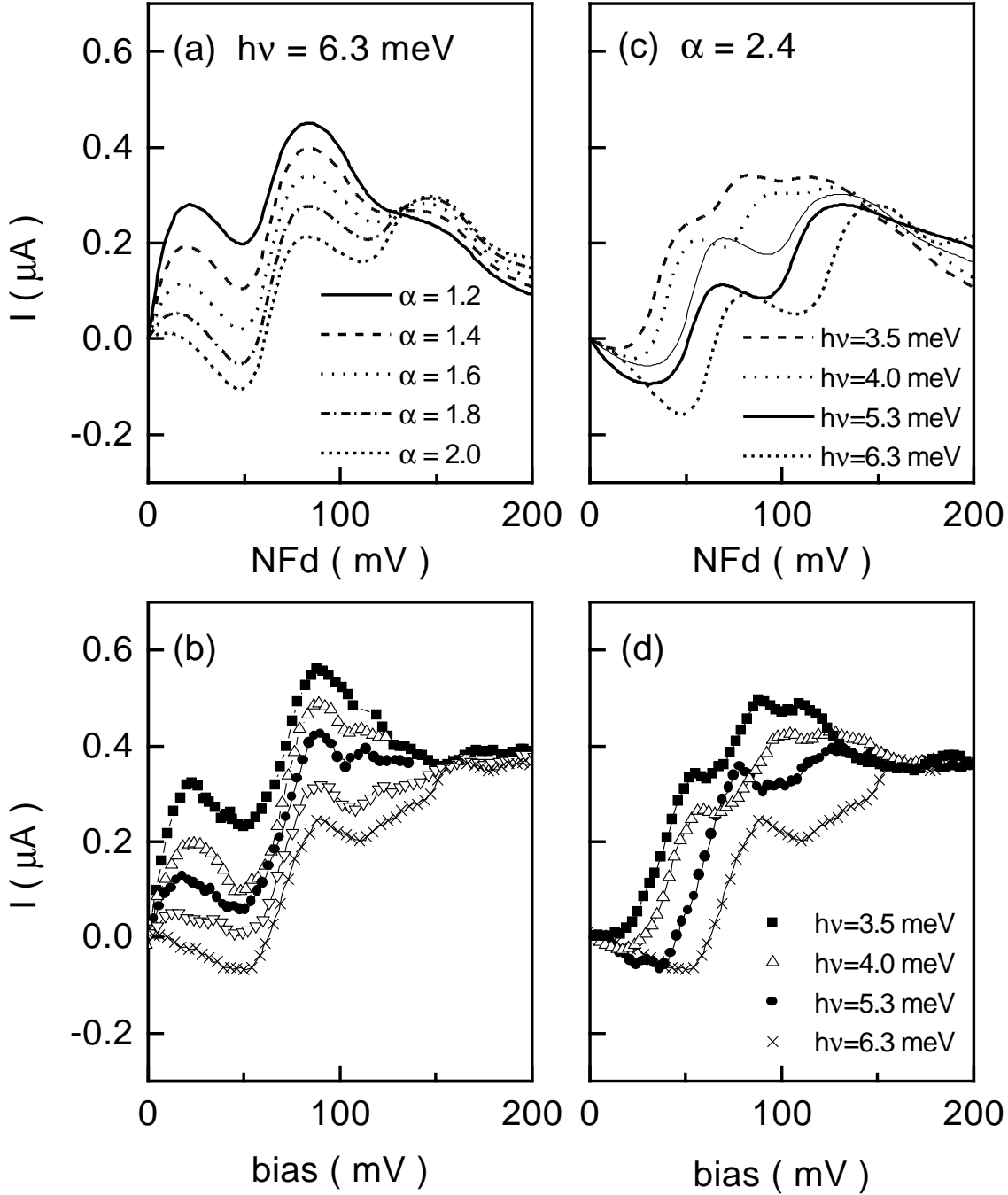}
\caption[a]{Current-voltage characteristics  under irradiation.
a) Theoretical results for $h\nu =6.3$ meV and different
field strength $eF_{ac}d=\alpha h\nu $ of the irradiation.
b) Experimental results for  $ h\nu=6.3$ meV and different
laser intensities increasing from the top to the bottom.
The actual  values $F_{ac}$ inside the sample are not accessible.
c) Theoretical results  for $\alpha=2.4$ and different
photon energies. The thin line depicts $h\nu=5.3$ meV and 
$\alpha=2.1$. d) Experimental results for different photon energies.
The laser intensity  was chosen to give maximum  negative conductance.}
\label{Figeinstrahl}
\end{figure}

Finally, we consider the sample in a strong terahertz field from a 
free-electron laser with 
frequency $\nu$ and field strength $F_{ac}$.
Following Refs.~\cite{KEA95a,ZEU96,PLAp} we use the relation
\begin{equation}
I_{n\to n+1}^{\rm irr}=\sum_{l=-\infty}^{\infty}[J_l(\alpha)]^2
I_{n\to n+1}(eF_{ac}d+lh\nu)\;,
\label{EqTucker}
\end{equation}
where $\alpha=eF_{ac}d/(h\nu)$ and $J_l$ is the ordinary Bessel
function of order $l$. This formula was originally  derived
for tunneling  through a single barrier\cite{TUC85} and holds
within a  miniband model for superlattices as well\cite{IGN95}. 
However, modifications due to   photon side-bands from a 
single quantum well\cite{WAG96} have been suggested. 
According to the arguments presented above, we model 
the nonequilibrium distribution function by assuming an 
electron temperature of 35 K. 
Then only 48 \% of the electrons are occupying the states below $E=0$.
Quantitative agreement between theory (Fig.~\ref{Figeinstrahl}a) 
and experiment (Fig.~\ref{Figeinstrahl}b) is found for
$h \nu=6.3$ meV (1.5 THz) for different strengths of the laser field 
\cite{Currents}.
We find a direct tunneling  peak at 
$U_{\rm dir}=N F_{\rm low}d\approx 20$ mV and photon replicas 
at $U\approx U_{\rm dir}+Nh\nu/e$ and 
$U\approx U_{\rm dir}+2Nh \nu/e$. For low bias and high intensities
there is a region of absolute negative conductance (NC)\cite{KEA95b}
which is studied for different frequencies in the following.
In the experiment at each frequency the laser power was chosen to give 
maximum NC near zero bias. As can be seen from 
Fig.~\ref{Figeinstrahl}(d) a minimum in current  occurs at 
$U\approx -U_{\rm dir}+ Nh\nu /e$. This is  just the photon replica 
of the direct tunneling peak on the negative bias side\cite{ZEU96} 
which dominates if the direct tunneling channel is 
suppressed close to the zero of $J_0(\alpha)$ in Eq.~(\ref{EqTucker}),
i.e., $\alpha\approx 2.4$, as used in the calculation 
(Fig.~\ref{Figeinstrahl}c). For $h \nu=5.3$ meV  a smaller value 
of $\alpha$ (thin line) agrees better with the experimental data.
This may be related to charge accumulation effects inside the sample
weakening the NC, so
that maximum NC is observed at a laser field corresponding to
a lower value of $\alpha$.
Both the theoretical and experimental result show that
absolute NC is persistent in a wide 
range of frequencies but becomes less pronounced 
with decreasing photon energy.

In conclusion, we have 
presented a self-contained theory of
transport in low doped and weakly coupled quantum wells.
A detailed comparison with several different
experiments shows that:
(i) it is necessary to consider impurity bands;
(ii) excellent agreement between theory and experiment is
achieved for static and irradiated IV as
well as zero-bias conductance using  reduced screening, and
(iii) the simple relation (\ref{EqTucker}) provides
a quantitative description of photon-assisted transport in multiple
quantum wells.


We want to thank Ben Hu for stimulating discussions.
A.W. and S.Z. acknowledge financial support by the
Deutsche Forschungsgemeinschaft.
Work performed at the Center for Terahertz Science and Technology was
supported by the Office of Naval Research, the Army Research Office
and the National Science Foundation.


\begin{thebibliography}{10}
\bibitem{Details}
We use $T_e=4K+26K\sqrt{eFd/8{\rm meV}}$ for $eFd<8$ meV and $T_e=30K$ for
  $eFd>8$ meV. The boundary conditions are $n_0=1.1N_D$, $n_{11}=N_D$ to model
  the charge accumulation at the injecting contact.
\bibitem{Currents}
Note that the calculated currents are slightly too large for the unirradiated
  case and slightly too small with irradiation. This discrepancy 
  may be related to deviations from the effective temperature model 
  used for the electron distribution functions.
\end{thebibliography}

\newpage

\newpage
\newpage

\newpage

\end{document}